\documentclass[epj,english]{webofc}
\usepackage[varg]{txfonts}   % Web of Conferences font
\usepackage[english]{babel}
\wocname{EPJ Web of Conferences}
\woctitle{INPC 2013}
\begin{document}
\title{Isospin transport in $^{84}$Kr+$^{112,124}$Sn reactions at Fermi energies}

\author{S.Piantelli\inst{1}\fnsep\thanks{\email{silvia.piantelli@fi.infn.it}} \and G.Casini\inst{1} \and A.Olmi\inst{1} \and S.Barlini\inst{2} \and M.Bini\inst{1,2} \and S.Carboni\inst{1,2} \and P.R.Maurenzig\inst{1,2}  \and G.Pasquali\inst{1,2} \and G.Poggi\inst{1,2} \and A.A.Stefanini\inst{1,2} \and R.Bougault\inst{3} \and N.LeNeindre\inst{3} \and O.Lopez\inst{3} \and M.Parlog\inst{3} \and E.Vient\inst{3} \and E.Bonnet\inst{4} \and A.Chbihi\inst{4} \and J.D.Frankland\inst{4} \and D.Gruyer\inst{4} \and E.Rosato\inst{5} \and G.Spadaccini\inst{5} \and M.Vigilante\inst{5} \and B.Borderie\inst{6} \and M.F.Rivet\inst{6} \and M.Bruno\inst{7} \and L.Morelli\inst{7} \and M.Cinausero\inst{8} \and M.Degerlier\inst{8}  \and F.Gramegna\inst{8} \and T.Marchi\inst{8} \and R.Alba\inst{9} \and C.Maiolino\inst{9} \and D.Santonocito\inst{9} \and T.Kozik\inst{10} \and T.Twarog\inst{10}
}

\institute{INFN - Sezione di Firenze, Italy 
\and
          Dip. di Fisica, Univ. di Firenze, Firenze, Italy 
\and
           LPC, IN2P3-CNRS, ENSICAEN and Univ. Caen, Caen, France
\and 
GANIL, CEA/DSM-CNRS/IN2P3, Caen, France
\and
INFN - Sezione di Napoli, Italy and  Dip. di Fisica, Univ. di Napoli Federico II, Napoli, Italy
\and
IPN - Orsay, Orsay, France
\and
INFN - Sezione di Bologna, Italy and Dip. di Fisica, Univ. di Bologna, Bologna, Italy 
\and
INFN - LNL, Italy
\and
INFN - LNS, Italy
\and
Jagellonian University, Institute of Nuclear Physics, Krakow, Poland
          }

\abstract{%
Isospin transport phenomena in dissipative heavy ion collisions have been investigated at Fermi energies with a beam of $^{84}$Kr at 35AMeV. A comparison of the $\langle N\rangle/Z$ of light and medium products forward-emitted in the centre of mass frame when the beam impinges on two different targets, the n-poor $^{112}$Sn and the n-rich $^{124}$Sn, is presented. Data were collected by means  of a three-layer telescope with very good performances in terms of mass identification (full isotopic resolution up to $Z\sim 20$ for ions punching through the first detector layer) built by the FAZIA Collaboration and located just beyond the grazing angle for both reactions. The $\langle N \rangle /Z$ of the products detected when the n-rich target is used is always higher than that associated to the n-poor one; since the detector was able to measure only fragments coming from the QuasiProjectile decay and/or neck emission, the observed behaviour can be ascribed to the isospin diffusion process, driven by the isospin gradient between QuasiProjectile and QuasiTarget. Moreover, for light fragments the $\langle N\rangle /Z$ as a function of the lab velocity of the fragment is observed to increase when we move from the QuasiProjectile velocity to the centre of mass (neck zone). This effect can be interpreted as an evidence of isospin drift driven by the density gradient between the QuasiProjectile zone (at normal density) and the more diluted neck zone.
 
}
\maketitle
\section{Introduction}
\label{intro}
The availability of detectors able to isotopically resolve the ejectiles of a heavy ion reaction at Fermi energies (30-50AMeV) allows to investigate the isospin degree of freedom and its evolution during the collision. Many experimental efforts were devoted to the study of the isospin diffusion process, by means of reactions involving partners with different N/Z (see e.g. \cite{tsang2004,liu,galichet,galichet2}); in fact the isospin diffusion is driven by the isospin gradient between projectile and target. Other studies (e.g. \cite{defilippo,mcintosh,lombardo}), sometimes involving symmetric reactions, were devoted to the investigation of the isospin drift i.e. the isospin enrichment of the low density neck zone with respect to the QuasiProjectile (QP) and QuasiTarget (QT) regions that are at normal density; this process is supposed to be driven by the density gradient between the two regions. As it is explained in many theoretical works \cite{ditoro,baran}, the study of these kind of phenomena is extremely important, because it can give information on the symmetry energy term in the nuclear equation of state. Recently also the FAZIA Collaboration \cite{fazia} has started to investigate these processes \cite{barlini}, thanks to the good capabilities in terms of isotopic resolution of the developed three-layer telescopes (full isotopic resolution up to $Z\sim 20-23$ for ions punching through the first detector layer). In this work we present some evidences of isospin diffusion and isospin drift observed by the FAZIA Collaboration for two reactions with the same beam, $^{84}$Kr at 35AMeV, and two different targets, the ``n-poor'' $^{112}$Sn and the ``n-rich'' $^{124}$Sn. The experiment was performed at LNS of INFN in Catania (Italy) by means of the beam delivered by the CS Superconducting Cyclotron. Data were collected by means of a telescope located just beyond the grazing angle for both reactions ($4.1^{\circ}$ for the n-poor target and $4.0^{\circ}$ for the n-rich one); the detector covered polar angles between $4.8^{\circ}$ and $6^{\circ}$. 

\section{Experimental results}
\label{sec-1}
\begin{figure}
\centering
\includegraphics[width=7cm,clip]{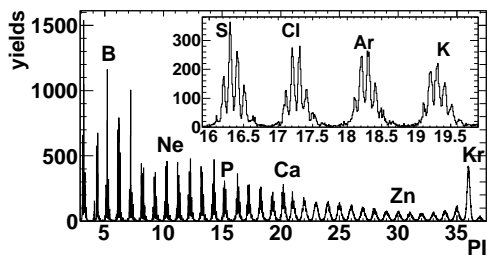}
\caption{Particle identification spectrum for ions punching through the first Si layer. In the inset the S - K region is expanded}
\label{fig1}       % Give a unique label
\end{figure}

The basic FAZIA detector consists of a three-layer telescope, with two reverse mounted n-TD Silicon detectors (thicknesses: 300$\mu$m and 500$\mu$m, respectively) manufactured by FBK, followed by a 10cm long CsI(Tl), manufactured by Amcrys and read out by a photodiode. The Silicon detectors have doping uniformity better than 3\% FWHM, thickness uniformity within 1$\mu$m and they have been obtained from wafers cut $7^{\circ}$ off the $\langle 100\rangle $ axis in order to minimize the channeling effect. The telescope is fully equipped with digital electronics. The identification procedure uses the standard $\Delta E$ - $E$ technique for particles punching through the first Si layer, obtaining full isotopic resolution in a region up to now accessible only by means of spectrometers (full isotopic identification up to $Z\sim 23$). For particles stopped in the first Si layer, full charge identification can be obtained by means of Pulse Shape Analysis techniques if the ions have a range in Si greater than 30$\mu$m (the threshold value increases with the ion charge). An example of the identification capabilities of the detector can be found in figure~\ref{fig1}, where the particle identification spectrum obtained by means of the $\Delta E$ - $E$ technique for ions punching through the first Si layer is presented. In particular, in the inset the zone between S and K is expanded and the isotopic resolution capability can be appreciated. More details on the detector performances and on the algorithms used by the Collaboration to treat the signals collected by the telescope can be found in \cite{carboni} and references cited therein. The results presented in this work refer only to particles punching through the first Si layer.

\begin{figure}
\centering
\includegraphics[width=6cm,clip]{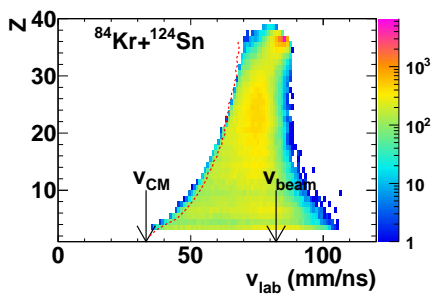}
\caption{Correlation between the charge and the lab velocity for ions detected by the FAZIA telescope for the n-rich reaction. The dashed line corresponds to the thresholds for the punch through of the first Si layer. The arrows correspond, respectively, to the center of mass velocity ($v_{CM}$) and to the beam velocity ($v_{beam}$). From \cite{barlini}}
\label{fig2}       % Give a unique label
\end{figure}

Since in this experiment we had only one telescope, only inclusive measurements could be done, without any selection on the event type. In any case, due to the geometrical position of the detector, some kind of selection is automatically done. In fact, as it is evident from the Z - lab velocity correlation reported in figure~\ref{fig2}, only fragments forward-emitted in the centre of mass can be detected by our telescope. Thanks to their position in the correlation of figure~\ref{fig2} and relying on the existing literature we can identify them as QP residues (for $Z\gtrsim 20$), QP fission fragments ($Z\sim 7-20$), neck emission (light fragments close to the centre of mass velocity) or QP evaporation (light fragments with velocity close to the beam velocity). The QT contamination is negligible.

\begin{figure}
\centering
\includegraphics[width=6cm,clip]{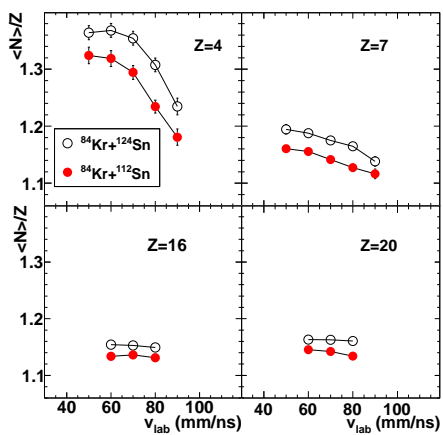}
\caption{$\langle N\rangle /Z$ as a function of the lab velocity. Each panel refers to a different element; open points correspond to the reaction $^{84}$Kr+$^{124}$Sn, while full points correspond to $^{84}$Kr+$^{112}$Sn.}
\label{fig3}       % Give a unique label
\end{figure}

Thanks to the good isotopic resolution of the FAZIA telescope, we are able to measure the average isospin for all the detected products with $Z \le 20$. The obtained average isotopic distribution as a function of the lab-velocity for both investigated systems is plotted in figure~\ref{fig3}, where open points correspond to the neutron rich case and full points to the neutron poor one. Each panel refers to a different element. For all ions and all lab-velocity values, the $\langle N\rangle /Z$ associated to the neutron rich system is higher than that found when the target is the neutron-poor $^{112}$Sn. For both reactions the same projectile ($^{84}$Kr) was used and the kinematics are very similar, as proved by the similarity of the grazing angles. Since the telescope can detect only fragments coming from the QP decay or from the forward (in the centre of mass frame) neck emission, the observed difference is an evidence of isospin diffusion, driven by the isospin gradient between target and projectile.  

Another observation emerging from this picture is the fact that while the $\langle N\rangle /Z$ for heavy products (bottom part) is independent of their lab-velocity, this is not true for light fragments (top part). In fact in this last case we observe a clear increase of the average isospin value when moving from velocities close to QP towards the centre of mass region. This last effect may be explained as an evidence of isopin drift: light fragments coming from neck emission (i.e. velocities close to the centre of mass) show a neutron enrichment with respect to those evaporated from the excited QP. According to theoretical models (e.g. \cite{ditoro}), this neutron enrichment is driven by the density gradient between the QP zone, where nuclear matter is at normal density, and the more diluted neck region. On the contrary, heavier fragments, which are supposed to come from the QP fission, do not show any velocity dependence of their average isospin content, because they come from a unique source (the fissioning QP).

The good capabilities of the FAZIA telescope have allowed us to extend the isotopic identification in a region which was up to now accessible only by means of spectrometers. As a consequence, further results on the isospin transport topic will be possible when, in the next future, more FAZIA telescopes will be available, allowing us to perform coincidences among the different ejectiles. In this way some selection on the centrality and on the event class will be possible in an extended solid angle with respect to what is accessible by means of spectrometers.

%
% BibTeX or Biber users please use (the style is already called in the class, ensure that the "woc.bst" style is in your local directory)
% \bibliography{name or your bibliography database}

\end{document}